\documentclass[prb,amsmath,amssymb,reprint]{revtex4-2}
\usepackage{amsmath}
\usepackage{graphicx}
\usepackage{bm}
\usepackage{xr}
\usepackage{todonotes}
\usepackage{enumitem}
\usepackage{xcolor}
\usepackage{hyperref}

\begin{document}

\newcommand{\red}[1]{\textcolor{red}{#1}}
\newcommand{\blue}[1]{\textcolor{blue}{#1}}

\title{Active $\Delta$-learning with universal potentials for global structure optimization}
\author{Joe Pitfield}
    \email{joepitfield@gmail.com}
\author{Mads-Peter Verner Christiansen}
\author{Bjørk Hammer}
    \email{hammer@phys.au.dk}
\affiliation{Center for Interstellar Catalysis, Department of Physics and Astronomy, 
    Aarhus University, DK-8000 Aarhus C, Denmak}

\date{\today}

\begin{abstract}
Universal machine learning interatomic potentials (uMLIPs)
have recently been formulated and shown to generalize
well. When applied out-of-sample, further data collection for
improvement of the uMLIPs may, however, be required. In this work we
demonstrate that, whenever the envisaged use of the MLIPs is global
optimization, the data acquisition can follow an active learning
scheme in which a gradually updated uMLIP directs the finding of new structures, which
are subsequently evaluated at the density functional theory (DFT) level. In the
scheme, we augment foundation models using a $\Delta$-model based on this new data using
local SOAP-descriptors, Gaussian kernels, and a sparse Gaussian Process
Regression model. We compare the efficacy of the approach with
different global optimization algorithms, Random Structure Search,
Basin Hopping, a Bayesian approach with competitive candidates (GOFEE),
and a replica exchange formulation (REX).  We further compare several
foundation models, CHGNet, MACE-MP0, and MACE-MPA. The test systems
are silver-sulfur clusters and sulfur-induced surface reconstructions on Ag(111) and Ag(100). Judged
by the fidelity of identifying global minima, active learning with
GPR-based $\Delta$-models appears to be a robust approach. Judged by the total CPU
time spent, the REX approach stands out as being the most efficient.
\end{abstract}

\maketitle

\section{Introduction}

Over the past two decades, the fields of molecular, colloid, and materials
science have supported the development of highly flexible machine
learning interatomic potentials (MLIPs)~\cite{Butler2018MachineLFA,
Lilienfeld2020RetrospectiveOAA,Fedik2022ExtendingMLA,D5CS00104H}.
Formulated using frameworks such as
feed-forward deep neural networks~\cite{Behler2007} or Gaussian Process Regression
models~\cite{Bartok2010}, such MLIPs have proven highly efficient for atomistic
simulations allowing, for e.g.\ larger and more complex systems~\cite{hormann2025machine} 
and longer time-scales
compared to simulations performed fully with density functional theory
(DFT)~\cite{C6SC05720A,MISHIN2021116980, PhysRevX.8.041048, Kapil_long_timescale}.
The data used to train such models is collected according to
various protocols~\cite{Maxson2024EnhancingTQA},
including random sampling~\cite{PhysRevLett.120.156001, PhysRevB.106.014102}, global optimization
searches~\cite{ronneAtomisticStructureSearch2022,Beacon2025}, normal mode
analysis~\cite{Tang2023}, molecular dynamics-driven sampling~\cite{KHORSHIDI2016310,timmermannMathrmIrOSurface2020},
and saddle point searching~\cite{Waters_2022}. Active learning schemes have been invoked, in which the
models are retrained upon assembly of data for which the models signal
uncertainty~\cite{zhi_pan_liu_active_learning_review_2019,PhysRevMaterials.3.023804,Bernstein2019,PhysRevB.100.014105,Bisbo2020,Behler2020,wangMAGUSMachineLearning2023,Kulichenko2024DataGFA,D4DD00231H,Alexandrova2025,venkat2025}.

Recently, equivariant graph-neural network models have been
introduced for MLIPs~\cite{pmlr-v139-satorras21a,batznerE3equivariantGraphNeural2022}.
These models encode the vectorial properties of
the local environment and embed information about farther environments
via message passing. Such models are generally more data efficient
~\cite{Leimeroth2025MachinelearningIPA} and
have supported the introduction of foundation models
~\cite{Chen2022AUGA,dengCHGNetPretrainedUniversal2023,batatia2023foundation},
where general purpose, "universal" MLIPs (uMLIPs) are trained on huge preassembled databases spanning the
entire Periodic Table and including compounds bonded covalently, ionically, dispersively, and as metals.
A swathe of recent advancements in the field of foundation models have 
shown rapid improvement in the benchmarks of these models and their suitability 
for direct application~\cite{taylor2025raffle,jakob2025universally}. 

A common procedure for making such advancements
and improving foundation MLIPs and MLIPs in general, focuses on
adding more diverse training data~\cite{deng2025systematic}. 
An example of this is the emergence of the successive models 
MACE-MP0 and
MACE-MPA~\cite{batatia2023foundation}. 
The training datasets are for both models that of the Materials Project database 
(MPtrj)~\cite{dengCHGNetPretrainedUniversal2023},
and in the case of MACE-MPA a portion of the Alexandria database~\cite{ghahremanpour2018alexandria}. 
The larger training dataset for MACE-MPA
causes the model to perform more accurately in predictions across the Matbench Discovery 
benchmark~\cite{riebesell2025framework}. An alternative approach for 
improving foundation MLIPs is via loss function engineering and change of design philosophy. 
The FAIR META eSEN-30M-OAM model~\cite{fu2025learning} shows iterative improvement by 
acknowledging the relevance of 
quality metrics, such as KSRME
(symmetric relative mean error in the phonon contribution 
to the thermal conductivity, which correlates with the smoothness of the PES) 
in describing when a potential is truly performative.  

Whilst it does not fall within the scope of most materials scientists to perform such 
bespoke model training from scratch (more so than ever with models such as 
the 1.6 billion parameter UMA model~\cite{wood2025family}), 
other methods of either changing or adding to existing models to fit a given 
application, do. When changing an existing model is limited in scope to adding more data (and potentially 
changing the loss function), the approach is commonly referred to as "fine-tuning". 

Fine-tuning has shown to be effective in improving the efficacy of models, so much so that 
the training methodology for some of the best performing models (eSEN) actually included 
fine-tuning explicitly. It has been found that such fine-tuning is effective at 
remedying systematic problems such as softening of the PES~\cite{deng2025systematic}. 
Other systematic examples, such as surface energies~\cite{focassio2024performance}, 
alloy mixing~\cite{marchand2025foundation}, thermal conductivity~\cite{pota2024thermal}, 
phonons~\cite{lee2025accelerating}
and sublimation enthalpies~\cite{kaur2025data}, have shown similar improvement 
under fine-tuning with 10s - 1000s of supplementary datapoints, depending on 
the property.


Crucially, the current metrics for fine-tuning or training large models in general do not 
place any emphasis on the energetic ordering of low energy structures. Often, the global minimum 
energy configuration for a system exists outside of the intuitive, thanks to effects which 
one can describe as phenomenological rather than systematic. uMLIPs, without proportional 
incentive to understand such phenomena, can become deficient in the discovery regime.
Studying silicate clusters and ultra-thin surface-oxides on Ag(111), we have previously demonstrated
that fine-tuning a uMLIP is reliably able to correct relatively minor structural inconsistencies and
yield correct ordering of low-energy configurations~\cite{pitfieldAugmentationUniversalPotentials2025}.
Notably, this required a dataset obtained through 
active learning enhanced exploration of the PES. This 
procedure would have been difficult and expensive to carry
out with iterative fine-tuning.

An alternative means to correct a uMLIP is that of an additive
$\Delta$-model~\cite{pitfieldAugmentationUniversalPotentials2025},
which is agnostic to several elements of the original
uMLIP, including i) the choice of loss function, ii) the network
architecture, and iii) the data used in the original training.
Such a correction is often able to encode phenomenological
effects, which are not well described by the uMLIP.  More so, where
such effects could cause catastrophic forgetting and deteriorate the
performance of the model, were they to be provided as small amounts of
training data for fine-tuning, a $\Delta$-model is less prone to
influence the interpolation domain. A further benefit of a scheme with
a $\Delta$-model is, that the training is computationally cheap and
can be applied often, not requiring the collection of large
batches of new data before updating the potential.

In the current work, we investigate the viability of establishing a
reliable $\Delta$-model corrected uMLIP via an active-learning
scheme. The use-case envisaged for the resulting corrected uMLIP is
global structure optimization, and hence the data-collection is guided
by structures realized during such optimizations. Different algorithms
are considered, Random Structure Search, Basin Hopping, and some more
elaborate ones introduced below. With these methods, the global minimum
DFT energy structures (GMs) of
a range of $\left[\mathrm{Ag}_2\mathrm{S}\right]_X$ clusters can be found via
active-learning of a $\Delta$-model added to a uMLIP. When using different uMLIPs
for active learning -- CHGNet, MACE-MP0, and MACE-MPA -- we find little
dependence on the rate of identifying the GMs
for the clusters, while for sulfur-induced reconstructions of silver surfaces some differences are
found, with MACE-MPA leading to the fastest discovery of the global
minimum energy structure. We will discuss this observation in terms of MACE-MPA
having encoded important Ag-S motifs more accurately, and hence
needing less data in the active-learning searches.

The paper is outlined as follows: In the methodology section we
introduce the three elements of the active learning: i) the
different uMLIPs used, ii) the Gaussian Process
Regression-based $\Delta$-model, and iii) and the four different global
optimization algorithms.  The results section starts with a discussion
of prior understanding of the uMLIPs without any correction. It proceeds by
presenting how the active learning performs for
$\left[\mathrm{Ag}_2\mathrm{S}\right]_X$ clusters when varying the
optimization method. Then the use of different uMLIPs 
in the context of active learning is considered,
first for the clusters and next for the sulfur induced surface
reconstructions. As the final topic in the results section, the use of
prior collected data for pre-correction of the uMLIPs is
considered. The paper ends with a discussion and details on data and code availability.

\section{Methodology}

\subsection{Universal MLIPs and training datasets}

MPTrj consists primarily of bulk structural 
information, namely 1.58 million unique structures and relaxation trajectories, 
computed at the PBE~\cite{perdew1996generalized} level of DFT. 
The Alexandria dataset is more diverse, containing 1D and 2D periodic systems 
alongside 3D bulk structures, providing more chemical insight into finite 
size effects particularly relevant to those investigated in this work. Overall, 
the dataset contains 30+ million structures, although 
this dataset is often sub-sampled to avoid structures being overrepresented 
(the sub-sampled dataset is \textit{only} 10 million structures). 
These structures are computed with the PBEsol~\cite{terentjev2018dispersion} and 
SCAN~\cite{sun2015strongly} XC functionals. 

In this work, we will examine three different uMLIPs namely CHGNet~\cite{dengCHGNetPretrainedUniversal2023}, 
MACE-MP0-large (henceforth referred to simply as MACE-MP0), and MACE-MPA-0 (MACE-MPA).
CHGNet is a graph neural network potential, in which site based magnetic moments
have been incorporated into the training data. The architecture of CHGNet involves 
both an atom graph where nodes are atoms carrying atomic embeddings and an auxiliary bond graph
where the nodes are bonds and edges carry angular information. Using interactions between 
these elements CHGNet incorporates angular information into the atomic embeddings. 
Ultimately, the atomic embeddings are used to predict total energies and magnetic moments 
in addition to forces and stresses through the use of automatic differentiation of the total energy. 

MACE is also a graph neural network, but unlike CHGNet it is equivariant with higher 
bond-order information gathered through tensor products involving directional information decomposed 
through spherical harmonics - making it capable of 
directly encoding angular information, dihedrals and beyond. 
These architectural differences have proven to improve performance 
across various benchmarks as evidenced by MatBench discovery~\cite{riebesell2025framework}.
We use two variants of MACE, namely MACE-MP0 trained solely on the MPTrj dataset that 
CHGNet is also trained on, and MACE-MPA which is 
trained on MPTrj and a subset of the Alexandria dataset. 
The larger training set used for the MACE-MPA uMLIP increases it performances 
on the aforementioned benchmarks compared to MACE-MP0. We employ the 'large' version 
of MACE-MP0, with $\sim 16$ million parameters and the medium version of MACE-MPA, with $\sim 9$ million parameters. 
Both models use the same building blocks and we attribute the majority of the differences in their behaviour that 
we observe to the difference in training set while acknowledging that we cannot prove this to be the case. 
In particular we later investigate the energetic ordering of clusters and observe significant differences in 
the ordering between MACE-MP0 and MACE-MPA, which we contribute to the training set differences but could also be 
partially caused by the architectural differences or even the random initialization of the network parameters. 

\subsection{$\Delta$-model}

Previously, $\Delta$-models have been used in bridging the gap between 
levels of theory~\cite{nandi2021delta}, including excited states~\cite{yang2024foundation} 
and prediction of correlation energies~\cite{bandyopadhyay2025accurate}. 
Their potential application in translating between levels of prediction 
has been demonstrated to be effective~\cite{ramakrishnan2015big}. Here, 
we apply that same translation between uMLIP predictions and DFT predictions using 
a $\Delta$-model~\cite{pitfieldAugmentationUniversalPotentials2025}. The corrected energy prediction for a structure, 
$\mathcal{M}$ (specifying super cell, atomic identities, and positions), is given by:
\begin{equation}
E_\mathrm{model}(\mathcal{M}) = E_{\mathrm{uMLIP}}(\mathcal{M}) + E_\Delta(\mathcal{M})
\label{eq:delta_model}
\end{equation}
where $E_{uMLIP}(\mathcal{M})$ is the unmodified uMLIP, and where
$E_\Delta(\mathcal{M})$ is the correction fitted to the residuals between the data and the 
incorrect uMLIP predictions.

In this work we evaluate
$E_\Delta(\mathcal{M})$ as a sparse Gaussian Process Regression 
based
on a Gaussian kernel evaluating the similarities between the local
atomic SOAP~\cite{bartokRepresentingChemicalEnvironments2013} 
descriptors of the query structure and the training
structures for the $\Delta$-model. See
Refs.~\onlinecite{ronneAtomisticStructureSearch2022} and \onlinecite{christiansen2024efficient}
for details on the GPR model.

Since the $\Delta$-correction in the present work is described by a
Gaussian Process (GPR), an interesting analogy exists, namely that
Eq.\ \ref{eq:delta_model} is equivalent to a GPR model in which the
uMLIP serves as a prior~\cite{lyngbyBayesianOptimizationAtomic2024,pitfieldAugmentationUniversalPotentials2025}.
The analogy, however, ceases to hold in cases where the uncertainty of the
uMLIP can be quantified and included in the GP, or when the $\Delta$-correction
is modeled with a neural network~\cite{christiansen2025delta}.

The training structures are comprised of the structures emerging from
the global optimization algorithms detailed in the next subsection. In
cases where data is available prior to the searches, some pretraining data will be included
in the pool of training data for the sparse-GPR model, while being excluded from the
pool of structures undergoing iterative improvement according to the global optimization scheme,
where relevant.

The SOAP descriptor is implemented in DScribe~\cite{dscribe,dscribe2}. 
We use $(n,l) = 3,2$ and
$r_{cut}=7$ {\AA}.  The sparse-GPR~\cite{ronneAtomisticStructureSearch2022} is implemented in AGOX~\cite{christiansen2022atomistic}. 
We use 1000 inducing points and train only on energies to keep the size of the
kernel matrix manageable.

\begin{figure*}[htb!]
    \includegraphics[width=1\linewidth]{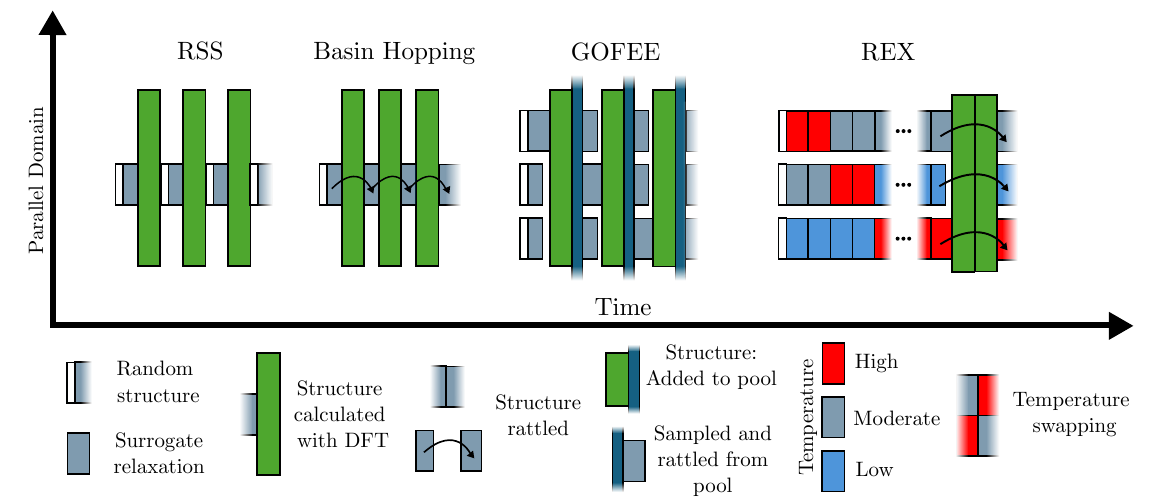}
    \caption{Schematic outlining the data collection and progression schemes 
    of RSS, Basin Hopping, GOFEE, and REX. The key highlights the processes 
    represented in the schematic. The conditioned acceptances,
    Eqs.\ \protect\ref{bh_metropolis} and \protect\ref{rex_metropolis}, of rattled and surrogate
    relaxed structures in Basin Hopping and REX are left out for
    clarity. The swapping events of REX are shown as walkers
    interchanging temperatures. In AGOX, this is coded as structures
    swapping structures.}
    \label{fig:ss_methods}
\end{figure*}

\subsection{Global optimization methods}
\label{go_methods}
Many choices exist for which global optimization method to employ
for the active learned data collection of the $\Delta$-model
introduced above. We have chosen a collection of methods ranging from
the almost completely unbiased random structure search, to a Monte Carlo
based Basin Hopping method, on towards more elaborate Bayesian- and parallel
tempering-inspired methods. This allows for an assessment of the
dependence of the active learning scheme on the optimization
method. All methods introduced will use the $\Delta$-corrected uMLIP
model for surrogate relaxation and perform the model update whenever
new DFT data becomes available. 
Relaxations are performed for up to a certain number of steps, 
or until a force convergence threshold is reached. For RSS and 
Basing Hopping, up to 500 relaxation steps are performed. For the Bayesian
method we use up to 100 relaxation steps, and for the parallel tempering-inspired method, 
30. The force convergence threshold in almost all cases is 0.05 eV/Å,
with the exception of the Ag(100)-$(\sqrt{17}\times \sqrt{17})$, for which 
a more stringent 0.025 eV/Å is used.
 Relaxations are conducted with the BFGS~\cite{fletcher2000practical} 
local optimization technique implemented in ASE~\cite{larsen2017atomic}.
Figure \ref{fig:ss_methods} presents a graphical overview of the global optimization methods used.
All of the optimization methods we employ are implemented in the modular framework of 
the AGOX global optimization package \cite{christiansen2022atomistic}.

Common to all
methods is that atoms are confined to \textit{confinement cells} both when
they are introduced and during relaxation. For each of 
the $\left[\mathrm{Ag}_2\mathrm{S}\right]_X$ 
structures, a $20\times 20\times 20 \ \mathrm{Å}$ unit cell 
houses a $15\times 15\times 15 \ \mathrm{Å}$ 
confinement cell. The sulfur-induced surface reconstruction searches are
done over 4 layer slabs of static Ag atoms in either Ag(111)-$(\sqrt{7}\times
\sqrt{7})$ or Ag(100)-$(\sqrt{17}\times \sqrt{17})$ cells using $(6\times 6)$
and $(2\times 2)$ $\mathbf{k}$-point grids for sampling of the corresponding 2D
Brillouin zones. The reconstructions are built from
$3\times$Ag$+3\times$S and $12\times$Ag$+8\times$S atoms introduced in
confinement cells with the same dimensionality as the lattices in the $xy$
directions and having a $z$ component of 4 {\AA}. All DFT calculations are performed
with plane wave GPAW~\cite{mortensen2024gpaw} (500 eV cutoff) 
and employing the PBE XC-functional~\cite{perdew1996generalized}.

\subsubsection{MLIP-assisted Random Structure Search (RSS)}

Random structure search is one of, if not \textit{the}, simplest structure search method 
widely applied in the history of computational materials science\cite{pickard2011ab}. 
Atoms are placed within
a cell at random whilst still accounting for reasonable bond lengths.
Some small constraints are placed
on the definition of random (no two atoms can be placed too close together, atoms must be 
placed within some distance of other atoms). In the present work, we require that 
no two atoms be placed closer together than 60$\%$ of the average of the covalent
radius of the two species, and that each atom must have at least one atom within
3 times this distance.

This structure, after surrogate relaxation, is then evaluated with first principles methods. 
As seen in Fig~\ref{fig:ss_methods}, each instance of RSS is entirely independent from others,
with the exception that the surrogate potential landscape is built from predecessor structures,
introducing slight correlation between structures over time.

\subsubsection{MLIP-assisted Basin Hopping (BH)}

Basin Hopping is a Monte Carlo based technique in which structural
walkers are evolved via random perturbations and relaxations~\cite{metropolis1949monte}.
In our implementation, all atoms are rattled according to a
uniform random distribution within a sphere around their
original position (whilst still accounting for reasonable bond lengths). 
The radius of the sphere is determined by the displacement magnitude,
which for Basin Hopping is set to a static 3 {\AA}. 
Structures are subsequently relaxed in the $\Delta$-corrected uMLIP. The new candidate, $\mathcal{M}_\text{new}$
is accepted according to the Metropolis Monte Carlo criterion probability, given by:
\begin{multline}
P_\text{accept}(\mathcal{M}_\text{new}|\mathcal{M}) \\
= \min\Big(1,\,
\exp\Big[
- \frac{E_\text{DFT}(\mathcal{M}_\text{new}) - E_\text{DFT}(\mathcal{M})}{k_B T}
\Big]\Big),
\label{bh_metropolis}
\end{multline}
where $E_\text{DFT}(\mathcal{M})$ is the energy of the current structure, $E_\text{DFT}(\mathcal{M}_\text{new})$  
is the energy of the proposed state, and the product of the Boltzmann constant and the 
temperature, $k_BT$, which is 0.15 eV. Basin Hopping introduces
explicit correlation in the time domain between structures considered --
i.e.\ structures depend on their structural predecessors, and
are commonly referred to as walkers to promote this notion. 
These are illustrated by arrows
on the Basin Hopping diagram shown in Fig~\ref{fig:ss_methods}. 
Regardless of the acceptance and propagation 
of the proposed structure through further iterations of the algorithm, 
its energy is calculated in DFT and added to the training set. 

\subsubsection{GOFEE}

Global Optimization for First-principles Energy Expressions (GOFEE) introduces
the concept that multiple walkers can be treated in parallel, as seen in 
Fig.\ \ref{fig:ss_methods}. GOFEE leverages principles of Bayesian statistics when creating new
structural candidates. It does so in two ways. Firstly, a set of 10 
structures are drawn from the pool of DFT evaluated structures. They 
are selected according to a K-means clustering of all structures in the pool, 
where each cluster contributes its lowest energy structure to a walker.
The walkers
are rattled as in Basin Hopping, but then relaxed in the lower
confidence bound,
$$
LCB(\mathcal{M}) = E_{model}(\mathcal{M}) -\kappa \sigma(\mathcal{M})
$$
where $\sigma(\mathcal{M})$ is the predicted uncertainty on $E_\text{model}$. $\kappa$ is a constant,
typically 2, which we maintain in this work.
The structure with the lowest value in the lower confidence bound
is then evaluated with DFT. This structure is finally added to the 
pool of evaluated structures, and a single DFT relaxation step performed. 
The process is then iteratively repeated. 
The process is then iteratively repeated. 
More details can be found in Ref.~\onlinecite{bisbo2022global}.
The uncertainty is evaluated employing an ensemble of GPR models trained on data with artificial
noise ($\sigma_p=0.001$ eV/atom, $\sigma_l=0.025$ eV/atom)
added as proposed in Ref.\ \onlinecite{christiansen2024efficient}.

\subsubsection{ML-assisted Replica Exchange X (REX)}

Like Basin Hopping, REX involves walkers which 
inherit the structure from the previous iteration, 
incrementally improving and exploring. 
Like GOFEE, it utilizes multiple instances of walkers progressing in 
parallel, sharing a surrogate potential between the walkers. 
This method also draws inspiration from replica
exchange (RE) methodologies in materials science
~\cite{sugita1999replica,zhou2002can,garcia2003folding,yamamoto2000replica,swendsen1986replica,unglert2025replica},
particularly parallel tempering (PT)~\cite{hansmann1997parallel,frantsuzov2005size}.
Parallel tempering is a special case of RE wherein the temperature is the 
only differing parameter between coupled replicas, and is often applied with
either Monte Carlo (MC) or molecular dynamics (MD) evolution, giving rise to the 
acronyms REMC and REMD, respectively.
These methods are both formulated to provide thermal samples for a given potential and overcome 
local minima and obtain ergodic behavior by maintaining an ensemble of walkers evolved in 
parallel at different temperatures. The $i$'th and $j$'th walkers are subject
to swapping events with probability:
\begin{multline}
\begin{aligned}
P_{\text{swap}}(i,j)
= \min\Bigg(1,\,
\exp\Big[ \big(&\frac{1}{k_BT_i} - \frac{1}{k_BT_j} \big) \\
\cdot \big(E_\text{model}(\mathcal{M}_j) - E_\text{model}(\mathcal{M}_i)\big) \Big] \Bigg),
\end{aligned}
\label{eq:swap_accept}
\end{multline}
which can be implemented either by swapping the candidates between fixed-temperature
walkers or by swapping the temperatures of the walkers that keep their structural candidates.
After an accepted swapping event, the involved structures
will propagate with altered temperatures. The rational is that in the
potential energy landscape high-temperature walkers identify new
regions of importance for sampling at lower temperature, and by making
these regions known to low-temperature walkers, the low-temperature
sampling converges faster. In a global optimization context, this
means that the GM will be identified more efficiently.

In our implementation, we leverage the main features of RE, but use Basin Hopping walkers,
i.e.\ walkers that are relaxed in the surrogate energy landscape of the
$\Delta$-corrected uMLIP before being subjected to the Metropolis
Monte Carlo acceptance test, which is further based on the surrogate energies,
\begin{multline}
P_\text{accept}(\mathcal{M}_\text{new}|\mathcal{M}) \\
= \min\Big(1,\,
\exp\Big[
- \frac{E_\text{model}(\mathcal{M}_\text{new}) - E_\text{model}(\mathcal{M})}{k_B T}
\Big]\Big).
\label{rex_metropolis}
\end{multline}
Due to the relaxations involved, detailed balance will be
violated, and hence rigorous thermal samples are not obtained by our RE
implementation. However, the method still benefits fully from the exploratory potential of coupled walkers. 

We dub our method Replica Exchange X (REX), where the X denotes the departure from previous methods, and 
the presence of other additions to the aforementioned algorithms, such 
as the uMLIP-based surrogate potential.

For each REX search, 10 walkers are instantiated with pseudo-random structures,
exactly as with RSS. Each walker is 
assigned a value for $k_BT$ according to a geometric series between 0.1 and
1. Similarly, each walker is assigned a different initial value for displacement magnitude,
linearly between 
0.1 and 5 Å. This determines the magnitude of 
attempted displacements, which are exactly as in Basin Hopping.
The highest temperature walker always generates a pseudo-random
structure.
The displacement magnitude
dynamically alters over the course of the calculation to target a $50\%$
metropolis acceptance rate for new structures. 
10 iterations of independent Basin Hopping events are performed between 
swapping attempts, and 30-50 between DFT calculations, where more complex 
systems are allowed more iterations between DFT calculations. When DFT is to be performed, 
5 structures are randomly selected from the 9 lowest temperature walkers thereby avoiding the
pseudo-random structure. 
This process is repeated until the allocated time budget has expired.

\subsection{$\left[\mathrm{Ag}_2\mathrm{S}\right]_X$ with uncorrected uMLIPs}
\begin{figure*}[htb]
    \includegraphics[width=1\linewidth]{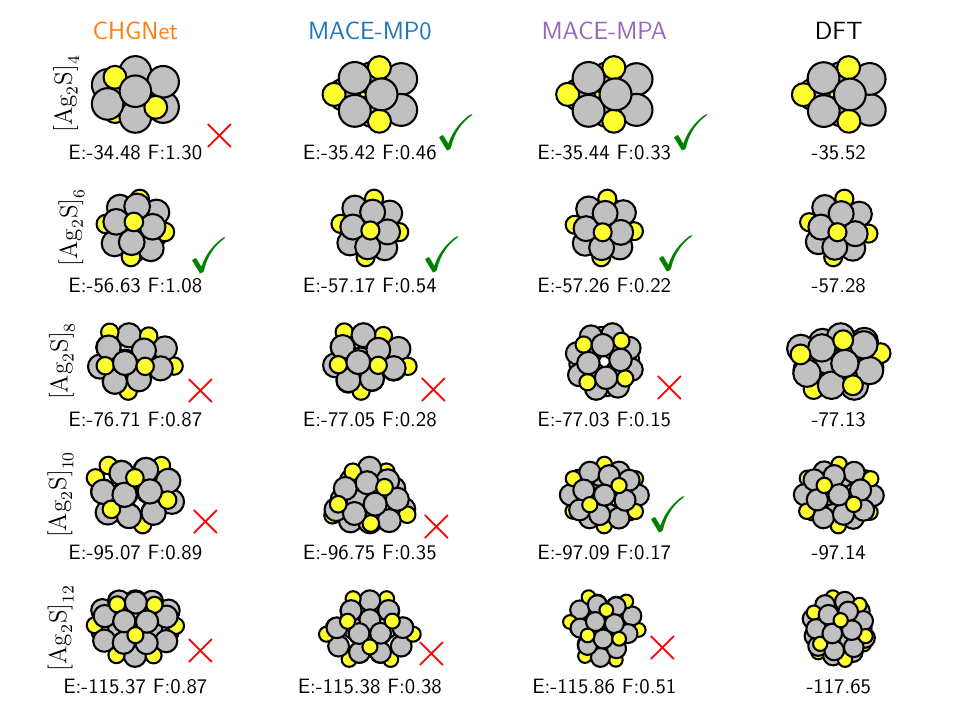}
    \caption{Global minimum energy structures in 
    CHGNet, MACE-MP0, MPA, and DFT. The energies (eV) and maximum atomic 
    forces (eV/\AA) calculated in DFT are shown underneath. 
    Green ticks and red crosses indicate if the structure identified by the universal
    potential matches configurationally with the DFT GM structure of 
    the corresponding stoichiometry.}
    \label{fig:cluster_predictions}
\end{figure*}

\section{Results}

The results of these methods when applied to a variety of 
systems is discussed here. 
We present these results by way of success 
curves~\cite{Bisbo2020,christiansen2022atomistic}.
Each vertical increment of a success curve indicates that one 
independent search has identified the solution at the 
given $x$-coordinate (usually CPU time or number of search episodes).
 Many searches together then 
provide a statistical ensemble of success (with each individual search
termed a repeat), which is 
more indifferent towards flukes. This can also be viewed
as the integral of the histogram of success. 
For the nanoclusters, 
we calculate the spectral decomposition graph for the best DFT structure in 
each search~\cite{christiansen2022atomistic}. Success is then
defined as having a graph which is identical
to that of the best structure.
In the case of surface reconstructions, we define success as 
having an energy within a threshold of the solution.
In order to obtain a 
comparable value for the CPU time, we ensure each search 
is performed on identical architecture and resources, namely 
10 cores from 48 of 2 Intel Xeon Gold 6248R CPUs, such that 
the CPU time is equivalent to ten times the walltime. DFT calculations are
run in parallel on all 10 CPUs. Surrogate calculations run on 1 CPU, and 
multiple will run in parallel across the 10 CPUs if allowed by the algorithm.

This section is organized as follows: We first discuss how well the
three uMLIPs describe the low-energy conformers of
$\left[\mathrm{Ag}_2\mathrm{S}\right]_X$ clusters for five values of
$X$. In more than half of the combinations of uMLIP and cluster size, it
is found that the global minimum uMLIP energy structure (uMLIP-GM) deviates
significantly from the global minimum DFT energy structure
(DFT-GM). The section proceeds to demonstrate that for a given uMLIP
the DFT-GM can be found using the proposed active learning $\Delta$-model.
From this analysis, it is further established that as the
cluster size and hence the complexity of the optimization problem
increases, more advanced search algorithms must be used. The section
moves on to consider the importance of the quality of the uMLIP. For
the $\left[\mathrm{Ag}_2\mathrm{S}\right]_X$ clusters, the
discrepancies between uMLIP-GMs and DFT-GMs do not appear to delay the
finding of the DFT-GMs in our active learning scheme. This suggests
that the problem of probing the proper configuration dominates the
problem of correcting the uMLIP. For sulfur-induced surface
reconstructions, the situation is reversed for the most complex problem, and MACE-MPA, which
needs less correction to describe the DFT-GM also facilitates its
faster finding. The section ends with a discussion of the usefulness
of precorrecting a uMLIP using data from prior searches.

Figure \ref{fig:cluster_predictions} shows the uMLIP-GM predictions of
the three universal potentials together with the DFT-GMs for
$\left[\mathrm{Ag}_2\mathrm{S}\right]_X$ clusters having
$X\in\left\{4, 6, 8, 10, 12\right\}$. The uMLIP-GMs were found by
conducting exhaustive REX searches, without the active
$\Delta$-learning, while the DFT-GMs were compiled from the results of
the active learning searches presented below in sections
\ref{active_learning_fixed_uMLIP} and \ref{active_learning_varying_uMLIP}. The cluster
illustrations are annotated with the total DFT energies and the maximum magnitude of 
the atomic forces. A green tick or red cross indicate whether or not
a DFT-based relaxation causes the structure to assume the
DFT-GM configuration, respectively.

For the smaller clusters, $X\in\left\{4, 6\right\}$, the uMLIP-GMs and
DFT-GMs agree except for the combination of CHGNet with
$\left[\mathrm{Ag}_2\mathrm{S}\right]_4$. Interestingly, this wrong
prediction involves a cluster of higher symmetry than the DFT-GM. In contrast, a
view at the DFT-GM for $\left[\mathrm{Ag}_2\mathrm{S}\right]_6$
reveals that it is a highly symmetric structure. This coincides 
with all three uMLIPs predicting the correct structure for this cluster
size, and hints that the network architectures and training datasets of the uMLIPs might favor high
symmetry.

For the larger clusters, $X\in\left\{8, 10, 12\right\}$, only the
MACE-MPA prediction for $\left[\mathrm{Ag}_2\mathrm{S}\right]_{10}$ is
correct, while all other predictions are wrong. We associate the
higher fail rate for the larger clusters with their configuration
spaces being considerably larger, putting the uMLIPs to a more stringent test.



Focussing on the DFT-based forces evaluated for the 
uMLIP-GMs, there is a clear tendency of decreasing force magnitudes
going from CHGNet, to MACE-MP0 and then MACE-MPA. Comparing
CHGNet and MACE-MP0, the difference must originate from the
network architecture, as they have been trained on the same materials
project dataset. Comparing MACE-MP0 and MACE-MPA, that are based
on more similar network architectures, the difference more likely lies in the
datasets, and it is seen that adding the Alexandria dataset has the
desired effect of leading to a more accurate uMLIP.

The DFT-GMs presented in Fig.\ \ref{fig:cluster_predictions} pertain
to a DFT setting using the PBE XC functional and the uMLIPs are
trained on PBE-based data, rendering the above comparison
meaningful. In the literature, GM structures for other XC functionals
have been reported. Using e.g.\ PBE0, a GM structure has been
reported~\cite{song2019systematic} which is similar to the MACE-MPA
prediction for the $\left[\mathrm{Ag}_2\mathrm{S}\right]_8$ cluster
shown in Fig.\ \ref{fig:cluster_predictions}. Relaxing our MACE-MPA
and DFT GM structures with PBE0 we confirm this result, but
we also find that by including the D3~\cite{D3} van der
Waals dispersion term and performing PBE0-D3 DFT calculations,
the PBE DFT-GM again becomes the preferred structure. We
stress, however, that since the currently used uMLIPs were trained on
PBE data, the fair comparison is to other PBE based results. Hence, the
green ticks in Fig.\ \ref{fig:cluster_predictions} indicate the uMLIP
predicting the correct PBE DFT-GM structure. It is also noteworthy 
that the $\Delta$-model strategy trivially extends to coupling 
uMLIPs trained on one level of theory to DFT calculations performed at another. 

\begin{figure}
    \includegraphics[width=\linewidth]{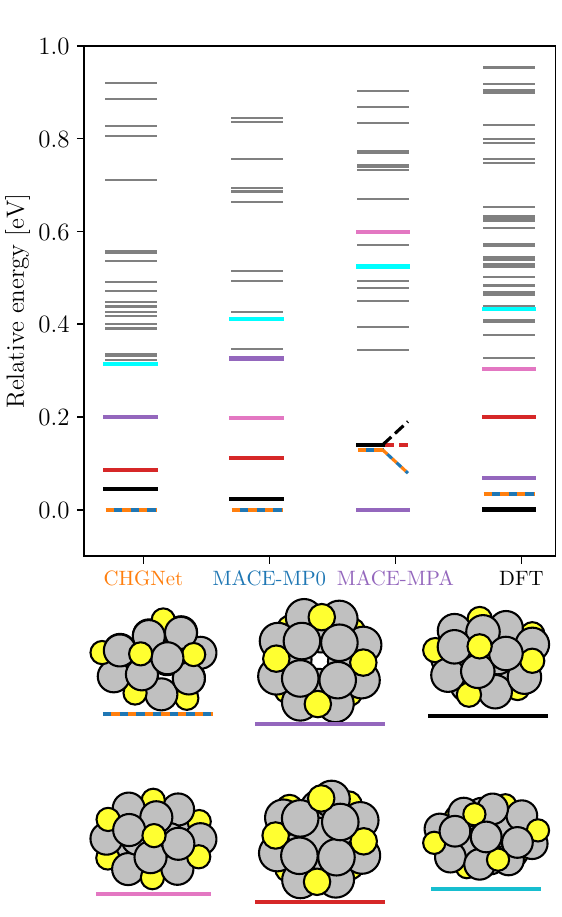}
    \caption{Depicted are the relative energies of a subsampling 
    of $\left[\mathrm{Ag}_2\mathrm{S}\right]_8$ structures relaxed in CHGNet, MACE-MP0, 
    MACE-MPA and DFT. These energies are relative to the lowest energy obtained 
    in any given potential. Structural diagrams are provided for a selection of 
    structures according to color.}
    \label{fig:Ag16S8_comparison}
\end{figure}

In order to provide a more comprehensive picture of quality of the
uncorrected uMLIPs, Fig.\ \ref{fig:Ag16S8_comparison} presents the
relative stability of a set of different low-energy conformers of the
$\left[\mathrm{Ag}_2\mathrm{S}\right]_8$ cluster. 
The structures are
the uMLIP-GMs (one of which is shared between CHGNet and MACE-MP0), the DFT-GM, two other 
structurally distinct nanoclusters (red, pink, cyan) and a distribution of further structures (grey). 

The structures in Fig.\ \ref{fig:Ag16S8_comparison} were obtained from a set of REX searches
either with active learning of a $\Delta$-model corrected MACE-MP0 model to obtain local DFT minima,
or without active learning in three sets of REX searches to identify local minima in each of the
three uMLIPs considered. The resulting dataset is thus representative for 
both DFT and the individual potentials. Overall, 10500 structures were obtained in this way. 
We subsample this dataset (for which $80\%$ of structures fall into one of four basins) to 
reduce repetition by selecting only 1 in 100 structures, resulting in 105 structures.
Each structure is then relaxed in each uMLIP or in DFT and filtered for similarity to 
other structures. The resulting structures thus represent the relative ordering of minima
as they appear in each potential.

Comparing the stability order of the various conformers within
each uMLIP to that in DFT, it is seen that the relative order is
highly sensitive to the model, as we demonstrate in Figure~\ref{fig:cluster_predictions}. 
Firstly, the CHGNet and MACE-MP0 
structures are configurationally alike, residing in the same basin and differing 
only through finer structural parameters, and thus
share a structural model and blue/orange coloring. 
The small differences between the MACE-MP0 and CHGNet structures
are nonetheless significant 
enough to contribute to a large difference between the respective energy 
predictions. As seen in Figure \ref{fig:cluster_predictions}, the CHGNet minimum 
structure has both high DFT forces and a higher energy than the similar structure 
suggested by MACE-MP0. Moreover, MACE-MPA undervalues this structure compared 
to its own suggested GM (purple) on the order 0.2 eV. When relaxed and evaluated
at the DFT level, this highly symmetric 
and well coordinated MACE-MPA-GM structure is slightly less stable 
than both the DFT-GM and the CHGNet/MACE-MP0-GM structures. On the contrary, when described
at the CHGNet and MACE-MP0 level, its stability is strongly underestimated compared to the
same GM structures.
Furthermore, MACE-MPA evaluates the DFT-GM, CHGNet/MACE-MP0-GM, and red structures as 
being almost energetically degenerate, a behaviour absent from the other potentials
and DFT. The MACE-MPA prediction for the pink and cyan structures is similarly worse than MACE-MP0. 

It seems as though the inclusion of the Alexandria dataset codes for the stability 
of such structures as the MACE-MPA-GM prediction, and highlights the importance of more 
diverse datasets in capturing 
the behavior of such phases. However, it is clear that adding more data (even data
which one might conclude is more relevant to the system in hand) 
is not guaranteed to improve or even maintain performance, with MACE-MPA having become 
more inconsistent on a majority of relative energy predictions than MPtrj only models. 

However, the impact of this larger dataset is the contrary 
when considering $\left[\mathrm{Ag}_2\mathrm{S}\right]_{10}$, for which 
MACE-MPA alone is able to reproduce the correct DFT-GM structure, 
cf.\ Fig.\ \ref{fig:cluster_predictions}.
There is demonstrable improvement in mean max
forces and the difference in energy between the potential minima 
and the DFT minima as the dataset size
increases, but also evidence that adding more data can change the 
performance in unexpected ways 
for systems which were previously consistent. Thus, it is a reasonable 
assumption to make that when 
applying such potentials to a problem, particularly one for which the ordering 
of low energy structures 
is relevant, one should employ corrective measures. 

As a final note, we return to the largest cluster considered in
Fig.\ \ref{fig:cluster_predictions}, the
$\left[\mathrm{Ag}_2\mathrm{S}\right]_{12}$. For this system the
performance of each uMLIP diverges and produce structures which
clearly obey some rationale of chemical understanding, but still
deviate to the order of at least 1 eV from the DFT-GM, which is a
highly symmetric combination of motifs observed in the smaller
clusters. In this limit, any of the selected
potentials would be unsuitable, and all would require correction.

\begin{figure*}[htb!]
    \includegraphics[width=1\linewidth]{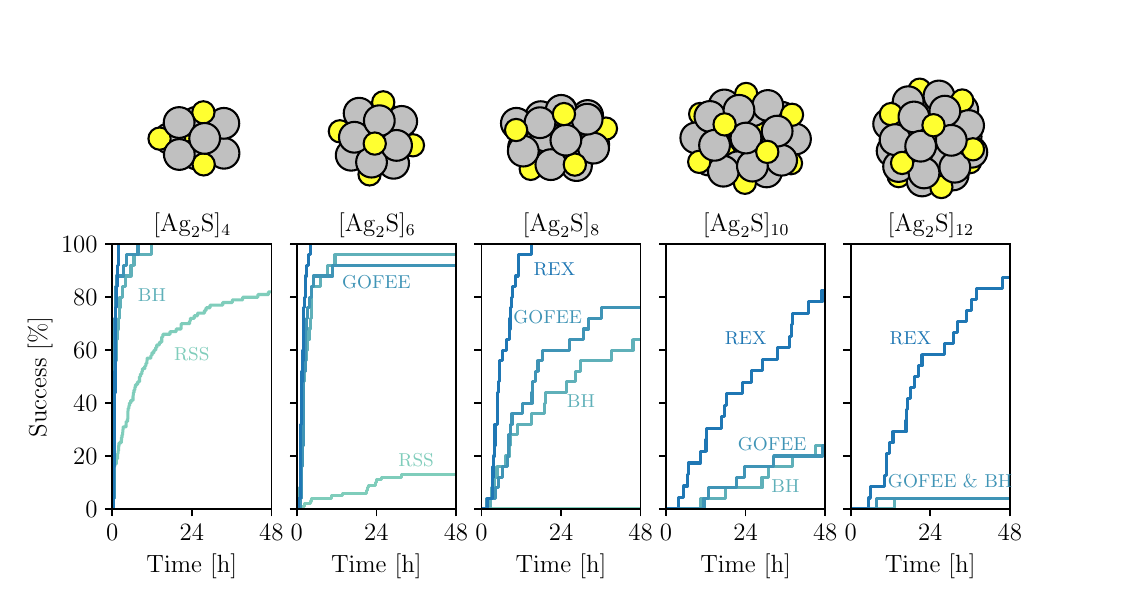}
    \caption{
      Global optimization of
      $\left[\mathrm{Ag}_2\mathrm{S}\right]_{X}$ clusters using active
      $\Delta$-learning with the MACE-MP0 universal potential and
      employing the four search methods outlined in Section
      \ref{go_methods}.  For each search 25 (100 for RSS) independent
      repeats were conducted and the success curves report the
      accumulated share of repeats that have found the GM as a function of elapsed time.
    The finding of the GM is determined according to a strict spectral graph decomposition.}
    \label{fig:method_comparison}
\end{figure*}

\subsection{Finding DFT-GMs from active learning of $\Delta$-model corrected uMLIP}
\label{active_learning_fixed_uMLIP}

In this section, we investigate the ease at which the DFT-GM can be
found while performing active learning to correct a uMLIP at the same
time as the global optimization is conducted. The five sizes of
$\left[\mathrm{Ag}_2\mathrm{S}\right]_{X}$ clusters present in
Fig.\ \ref{fig:cluster_predictions} are considered, and for each of
them, the different search algorithms presented in section
\ref{go_methods} are employed. For the uMLIP, MACE-MP0 is used
throughout, and the discussion of how different uMLIPs perform is
deferred to section \ref{active_learning_varying_uMLIP}. For the
smaller clusters, MACE-MP0 already codes for the correct DFT-GM
and requires no $\Delta$-model correction, but we include these cases for
completeness.

Figure \ref{fig:method_comparison} compares the four different structure search methodologies 
previously outlined, when applied to the five $\left[\mathrm{Ag}_2\mathrm{S}\right]_{X}$ cluster sizes. 
Firstly, $\left[\mathrm{Ag}_2\mathrm{S}\right]_4$ is a very simple problem to solve configurationally,
which can be understood intuitively from 
the relatively small number of atoms present. Furthermore, MACE-MP0 
encodes sufficient information to solve the problem correctly (see Fig.\ \ref{fig:cluster_predictions}),
so the impact of 
active learning on the outcome is likely minimal. Figure~\ref{fig:method_comparison} 
evidences this, with all but RSS solving the problem almost immediately. Even in this 
very simple regime, RSS is far from matching the other methods, succeeding in only 80$\%$
of cases after a substantial investment of resources. 

The struggle of RSS continues for $\left[\mathrm{Ag}_2\mathrm{S}\right]_6$, where once 
again MACE-MP0 is able to predict the solution 
correctly itself (cf.\ Fig.\ \ref{fig:cluster_predictions}). Here, RSS has a $\sim$15$\%$ chance of 
finding the solution given 48 hours of calculation time (480 CPU hours), 
compared to the near 100$\%$ chance in a fraction of the time for the other methods. 
For this cluster size, it further becomes apparent that 
the proposed REX methodology outperforms both GOFEE and Basin Hopping. 

$\left[\mathrm{Ag}_2\mathrm{S}\right]_8$ is the point where the configurational complexity 
and the incomplete map of the PES provided
by the uMLIP begin to differentiate the approaches. 
RSS fails to uncover the true structure even once across all repeats, 
indicating that attempting relaxation of almost purely random structures 
as a strategy for identifying the minimum of the 
PES whilst simultaneously correcting a surrogate 
potential becomes an unreliable strategy quite rapidly. This highlights 
the importance of 
the search strategy, and demonstrates
that configurational complexity correlates with the system size, when 
comparing $\left[\mathrm{Ag}_2\mathrm{S}\right]_X$ with $X\in\left\{4, 6, 8\right\}$.
RSS is omitted from further examples with $X >{8}$ on these grounds. 
REX, however, demonstrates $100 \%$ success in about one third of the 48 hours provided, 
thereby outperforming
GOFEE and Basin Hopping, which reach $\sim 75\%$ and $\sim 65\%$ in the full duration, respectively.

Similarly for $\left[\mathrm{Ag}_2\mathrm{S}\right]_{10}$, the performance of GOFEE and Basin 
Hopping drop with respect to REX, which while slower than for the smaller stoichiometry still 
achieves a success of 80$\%$ in the allotted time.

The same trend is echoed once again for $\left[\mathrm{Ag}_2\mathrm{S}\right]_{12}$,
with diminishing success for the GOFEE and Basin Hopping, and $>80\%$ for REX. 
Now, only one in 25 Basin Hopping or GOFEE repeats are able to obtain success at this size.
The single early success suggests a coincidental occurrence of the solution, indicating 
that the exploration becomes stunted as time progresses.
Meanwhile, 
REX continues to perform without the same abrupt decrease in success. 
This demonstrates that REX is a powerful explorative tool, effective at 
increasing systems sizes and  
regardless of the understanding of the underlying universal potential. 
REX effectively couples augmentative active learning to configurational exploration.

Notably, 
in the regime where \textit{ab initio} evaluation dominates the computational resource budget 
(many atoms, for periodic systems many $\mathbf{k}$-points, expensive XC functionals, and wavefunction methods),
REX is at a further advantage 
when compared to the other methods. Both Basin Hopping and GOFEE devote a larger fraction of 
their compute budget to performing DFT, and thus perform more DFT calculations per unit time than REX. 
This can be seen in Supplementary Figure 1, which
depicts the same data as Figure 
\ref{fig:method_comparison}, repostulated in terms of the number of first-principles calculations performed.
This alternative metric more heavily favours REX, as the number of DFT calculations is agnostic to the extent of the
repeated 
replica exchange cycle, and is indicative that REX would also perform excellently in the 
limit that the computational cost of first principles evaluation is the dominant factor.

\subsection{Comparing the success of different uMLIPs}
\label{active_learning_varying_uMLIP}
\begin{figure*}
    \includegraphics[width=1\linewidth]{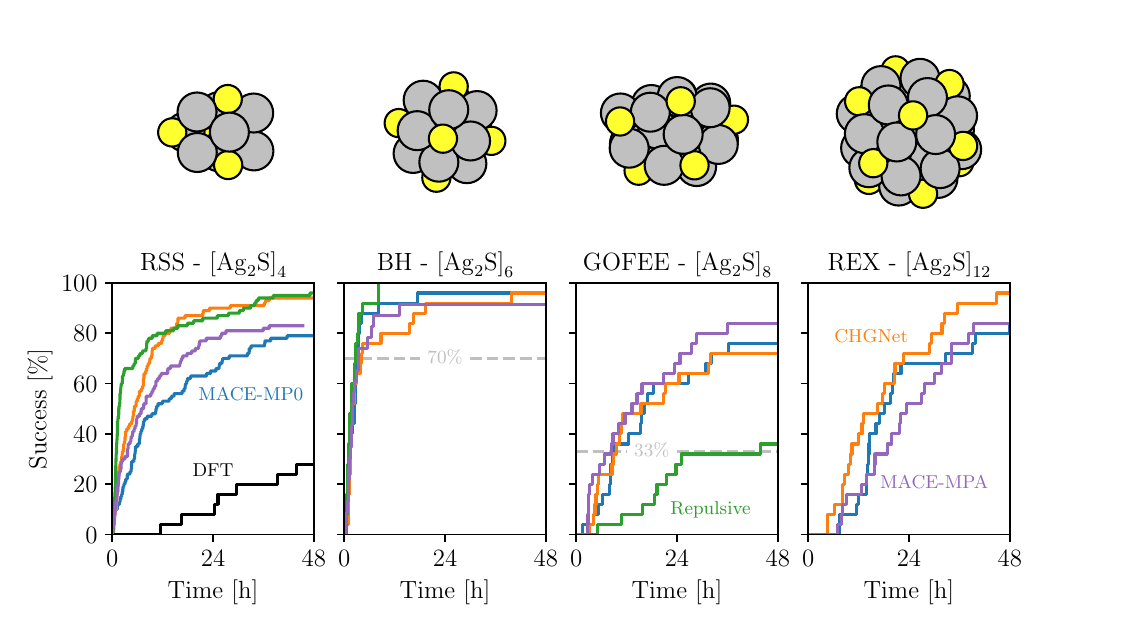}
    \caption{Success curves for global optimization of
      $\left[\mathrm{Ag}_2\mathrm{S}\right]_{X}$ using various
      combinations of optimization method (left to right) and uMLIP
      potentials (orange, blue, purple). Included are results using no
      uMLIP but only a repulsive prior (green) and omitting a surrogate potential altogether (black).
    }
      \label{fig:compare_potential_priors}
\end{figure*}

The previous section demonstrated for MACE-MP0 that the
DFT-GMs can reliably be determined in an active learning
$\Delta$-model approach. In this section, we widen the
investigation to the two other uMLIPs introduced. Since we have already
seen the limitations of the optimization algorithms we limit the study
to a few combinations of algorithms and cluster sizes.

Figure \ref{fig:compare_potential_priors} presents the
results. Starting with the RSS algorithm, the smallest cluster,
$\left[\mathrm{Ag}_2\mathrm{S}\right]_{4}$, is considered. All three
uMLIPs support the finding of the DFT-GM very well with this
method. In particular, the CHGNet, which has the wrong GM encoded,
cf.\ Fig.\ \ref{fig:cluster_predictions}, appears efficient, which we
attribute to the fast inference times of that uMLIP.

That speed is more
important than precise insight becomes apparent when omitting 
the uMLIP and instead using a simple
repulsive prior~\cite{bisbo2022global} that only acts to avoid the collapse of
atoms onto the same positions. The success curve of RSS with a
repulsive prior is included in green in
Fig.\ \ref{fig:compare_potential_priors} and is seen to reach success
in e.g.\ half of the repeats in a matter of 1-2 hours (10-20 CPU hours),
for which the uMLIP assisted methods require 5-10 hours (50-100 CPU hours). 
That a simple potential is useful for solving such problems is consistent
with findings of other investigations~\cite{PhysRevB.106.014102}.

Another testimony to the importance of speed of the underlying
relaxation calculations comes from the black curve in
RSS Fig.\ \ref{fig:compare_potential_priors}. The curve shows the success
obtained from a purely DFT-based RSS search for the
$\left[\mathrm{Ag}_2\mathrm{S}\right]_{4}$ cluster. This search has no
uMLIP and does not construct a surrogate potential to enable cheap
structural relaxations, rather every relaxation step is done at the 
full DFT level. This example represents a historical benchmark 
moreso than a viable modern strategy.

The Basin Hopping algorithm is applied to the 
$\left[\mathrm{Ag}_2\mathrm{S}\right]_{6}$ nanocluster. The rate 
at which each tested potential reaches 70$\%$ success is 
indistinguishable, with small deviations occurring past this point.
Interestingly, the simple repulsive potential is the only one to reach
100$\%$ during the allocated time, which as for 
$\left[\mathrm{Ag}_2\mathrm{S}\right]_{4}$ highlights the importance 
of the inference time of the surrogate potential for such 
problems.

The next optimization algorithm considered in
Fig.\ \ref{fig:compare_potential_priors} is GOFEE, which is tested on
$\left[\mathrm{Ag}_2\mathrm{S}\right]_{8}$. The rate at which the
DFT-GM is identified starts out very similar for the three
uMLIPs. After about 10 hours (100 CPU hours)
one in three repeats have indeed
found the DFT-GM, irrespective of which uMLIP is used. 
From there on, the performance of the uMLIPs remains more or less
consistent, with MACE-MPA performing the best, reaching over $80\%$ success
in the allotted time.

Like for RSS and Basin Hopping, GOFEE can be run without a uMLIP, using instead
a repulsive prior in the surrogate potential as originally conceived~\cite{Bisbo2020}.
This results in  the green curve for
$\left[\mathrm{Ag}_2\mathrm{S}\right]_{8}$ in
Fig.\ \ref{fig:compare_potential_priors}, but unlike for RSS and Basin Hopping, this
is now far less efficient than when a uMLIP is available, and
requires the full 48 hours (480 CPU hours) to just about reach the 33 \% success
rate. That the use of the uMLIP is more efficient, we attribute to the increased
complexity of the energy landscape for the $\left[\mathrm{Ag}_2\mathrm{S}\right]_{8}$ compared
to those for the smaller clusters solved with RSS and Basin Hopping, and we conjecture
that despite its faster inference time, the repulsive prior-based potential suffers
from requiring more datapoints for a reliable description of the energy landscape.

Figure \ref{fig:compare_potential_priors} finally presents the results
of search for the DFT-GM with the REX method for the
$\left[\mathrm{Ag}_2\mathrm{S}\right]_{12}$ cluster.
As the REX method relies heavily on many
rattle-relaxation cycles of walkers, it benefits from the use of
the faster CHGNet method when evaluated with a time comparison
metric. Notably, even at this size, REX obtains a high success rate 
of $>80\%$ regardless of the choice of uMLIP.

We note that it is not possible to use the REX method just based on a
repulsive prior, as the extensive searches done by the REX method
before any $\Delta$-model can be established would find that atoms
should separate as far as possible. This could be circumvented by
using a prior with some atomic attraction or by applying pretraining
 based on precalculated data.

Summarizing this section, we find surprisingly little variance in
the efficacy of the three different uMLIPs considered. Despite their
differences in initial accuracy, they all support the usage as initial
surrogate models and lend themselves to being corrected via our active
learning $\Delta$-model protocol.

\subsection{AgS Surface reconstructions}

\begin{figure}[htb]
    \includegraphics[width=0.75\linewidth, trim={8cm 2cm 8cm 2.2cm},clip]{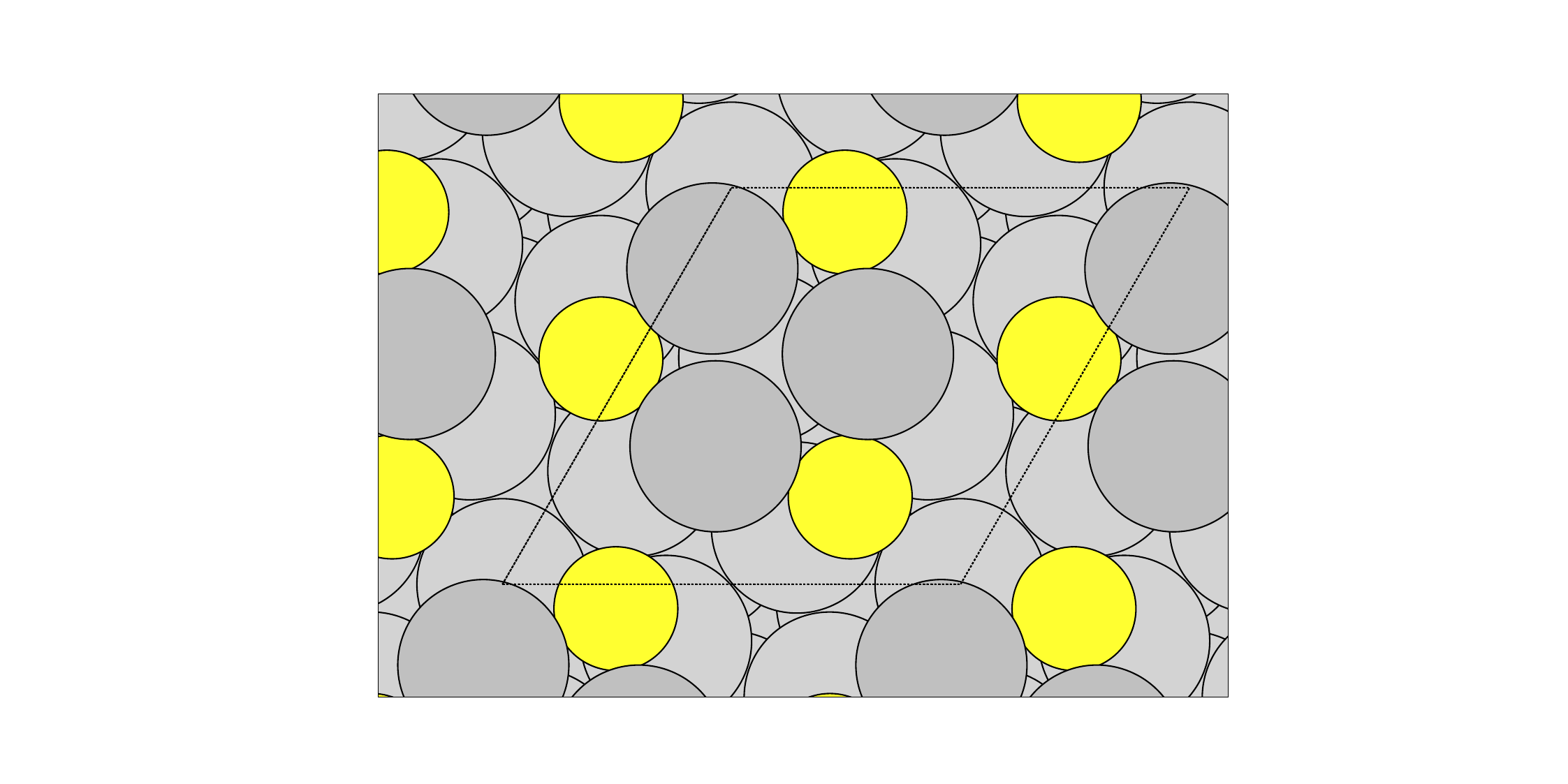}
    \caption{Depicted is the DFT GM structure for the $(\sqrt{7}\times\sqrt{7})$
    surface reconstruction of Ag(111) under sulfurization. Silver atoms 
    which do not reorganize have their color lightened to highlight the 
    reconstruction.}
    \label{fig:sqrt7sqrt7}
\end{figure}

\begin{figure*}[htb!]
    \includegraphics[width=1\linewidth]{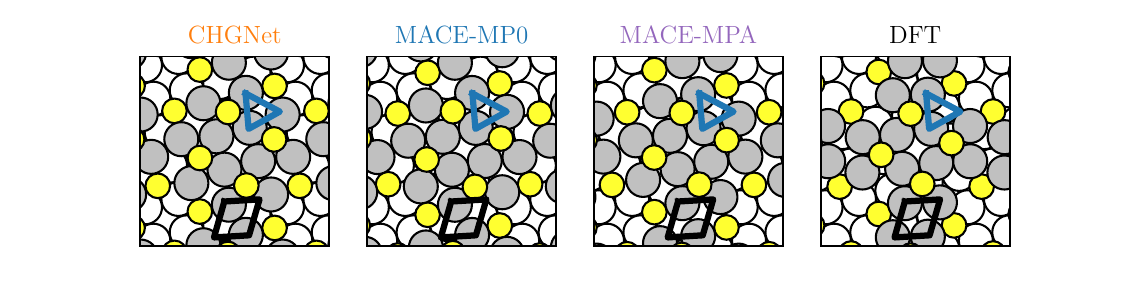}
    \caption{Structural diagrams of the sulfur-induced surface reconstruction 
    in the Ag(100)-$(\sqrt{17}\times\sqrt{17})$ unit cell. From left to right, 
    the minima are obtained through REX searches in CHGNet, MACE-MP0, 
    and MACE-MPA, without active learning. The DFT GM is obtained through REX searches 
    with MACE-MP0 
    as a prior. The black rhombus present in each panel shows the positions of four particular 
    silver atoms in the DFT solution. The blue triangle highlights three atoms in the MACE-MP0 solution.}
    \label{fig:surface_structures}
\end{figure*}

We now move to consider sulfur-induced surface reconstructions.
Based on the experimental information available~\cite{shen2008novel,russell2011adsorption},
the two systems, Ag(111)$-(\sqrt{7}\times\sqrt{7})$-Ag$_3$S$_3$ and
Ag(100)$-(\sqrt{17}\times\sqrt{17})$-Ag$_{12}$S$_8$, were chosen. With
their different sizes and hence complexity of the involved GMs, these
two systems allow for an assessment of the efficiency of the presented
methods for active learning and global optimization.  The GM structure
for the two systems are shown in Figs.\ \ref{fig:sqrt7sqrt7} and
\ref{fig:surface_structures}, respectively. For the smaller system,
all three uMLIPs and the full DFT description agree on a GM in which a
triangular Ag$_3$S$_3$ motif forms atop the Ag(111). This motif is
reminiscent of some of the facets found on the
$\left[\mathrm{Ag}_2\mathrm{S}\right]_X$ clusters.
The structure shown as the DFT GM in Fig.\ \ref{fig:sqrt7sqrt7}
agrees fully with that proposed in Ref.\ \onlinecite{shen2008novel}. Rotating
the Ag$_3$S$_3$ motif by 60 degrees brings the Ag atoms from \textsc{fcc} to \textsc{hcp} sites, and is associated
with a small $\sim$0.01 eV energy penalty. In our statistical
analysis below, we therefore do not discriminate between these two
solutions.

For the larger system, none of the uMLIPs predict the correct GM as
given by DFT. This is detailed in Figure \ref{fig:surface_structures} from
which it appears that CHGNet and MACE-MP0 energetically
overvalue the formation of the Ag$_3$S$_3$ motif
indicated by the blue triangle (the presence of this motif is
unsurprising, given its prevalence in examples thus far).  This
overvaluing draws the required silver atom away from the slanted
Ag$_4$S$_4$ motif indicated by the black rhombus.  MACE-MPA appears
aware that forming the triangles is not as much of an energetic
priority, but as seen in the black rhombus, still fails to correctly
establish the Ag$_4$S$_4$ motif.  This failure is reflected by the
slight clockwise rotation of the structure with respect to the surface
present in the DFT solution i.e.\ silver atoms in the reconstruction do
not fill exact sites of the lattice (hollow),
but rather shifted sites.
Comparing the DFT-GM of Fig.\ \ref{fig:surface_structures} to
that suggested in Ref.\ \onlinecite{russell2011adsorption} leads us to believe
that the present search has revealed a new, more stable PBE DFT-GM
than hitherto proposed, which further testifies to the efficiency of
the REX method in combination with active learning on top of a
uMLIP.

Clearly each of the tested uMLIPs has remarkable understanding of the general chemistry involved, 
and the nature of the required correction, whilst variable between potentials, is small. 

\begin{figure*}
    \includegraphics[width=1\linewidth, trim={0cm 1.2cm 0cm 0cm},clip]{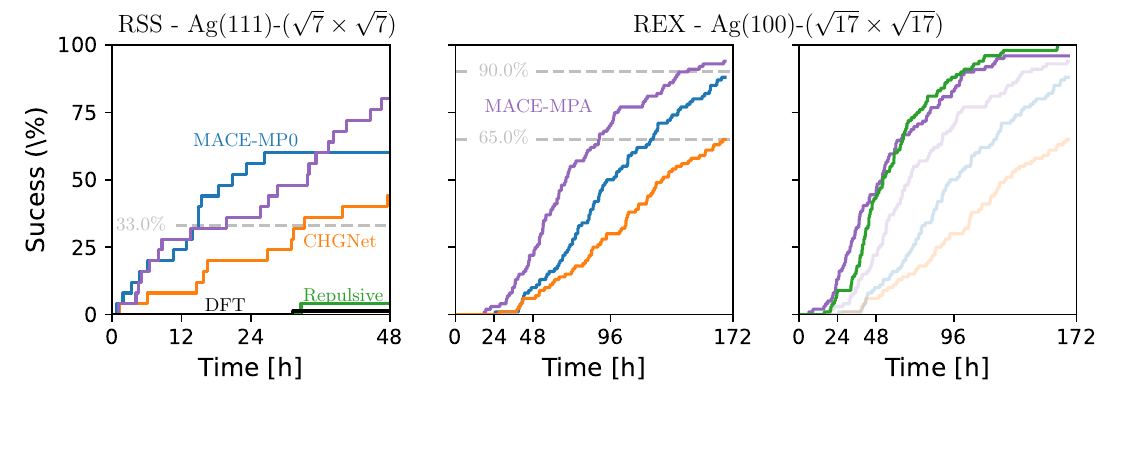}
    \caption{Success curves for (left) the RSS searches for
      Ag(111)$-(\sqrt{7}\times\sqrt{7})$-Ag$_3$S$_3$ and (middle, right) for the
      REX searches for Ag(100)$-(\sqrt{17}\times\sqrt{17})$-Ag$_{12}$S$_8$. The curves
      are colored orange, blue, and purple according to the universal potential used,
      green for repulsive prior instead of a universal potential, and black for DFT-only omitting
      the surrogate potential altogether. In the right panel, results are shown for
    two pretrained potentials: a $\Delta$-model corrected MACE-MPA,
    which has been provided with initial data (shown in purple), and a 
    GPR model with a simple repulsive prior which has been provided
    the same data (shown in green). The curves from the previous panel are
    translucently overlaid for reference.
    Success is evaluated according to an energy threshold.}
    \label{fig:surface_structures_search}
\end{figure*}

Figure \ref{fig:surface_structures_search} presents the results when
the GMs for the surface reconstructions are searched with active
learning using the different uMLIPs. The first panel of
Fig.\ \ref{fig:surface_structures_search} shows the case of the small,
Ag(111)$-(\sqrt{7}\times\sqrt{7})$-Ag$_3$S$_3$ problem, where the
uMLIP all code for the right GM. 
Owing to the smallness of the problem, we employ the
simple RSS optimization method. 
The success curves show that the GM
may be found with 33 \% success, i.e.\ in every 3rd repeat, after about 
14 hours (140 CPU hours)
in the two MACE potentials, 
with a considerably slower timeframe for CHGNet. 
Contrasting to this success, 
both a set of 75 full DFT random structure searches,
and 25 repulsive prior GPR enhanced RSS searches, 
obtained the solution to the problem only once each in the provided time.
We attribute 
this substantial decrease in performance to the increase in 
cost of DFT for surface problem when compared to nanoclusters. 
Clearly the uMLIP $\Delta$-model enhanced RSS 
enabled this search to succeed where
more traditionally it would have been very expensive.

The active learning results for the larger system of
Ag(100)$-(\sqrt{17}\times\sqrt{17})$-Ag$_{12}$S$_8$ are shown in the
second panel of Fig.\ \ref{fig:surface_structures_search}.  Due to the
increased complexity of the problem, the REX optimization method is
used. Despite the uMLIPs being confused about the exact GM
(cf.\ Fig.\ \ref{fig:surface_structures}), the correct DFT GM is
eventually found reliably with all three uMLIPs using the $\Delta$-model
approach.

For this system reasonable differences in the efficacy of
using the various uMLIPs are found. CHGNet reaches $65\%$ success
in 172 hours (1720 CPU hours), whereas the two MACE models 
reach around $90\%$ in the same timeframe. 

MACE-MPA is found to be considerably faster than the other 
two uMLIPS at obtaining the solution in general, requiring just 
over 90 hours to reach the same success as CHGNet did in the full 172. 
That MACE-MPA is the one to
perform the best can be rationalized in terms of its GM deviating less
from the DFT-GM than the other two uMLIPs,
as discussed in connection with\ Fig.\ \ref{fig:surface_structures}. 
The $\Delta$-model thus
requires less data to be gathered by the active learning in order to
correct the uMLIP.

Unlike the nanocluster example, none of the uMLIPs fully encode for the 
solution. This is not particularly surprising, given that the 
difference between the DFT solution and the uMLIP solutions boils 
down to a small rotation of the reconstruction with respect to the 
underlying surface layer, a highly specific surface phenomenon.

To create such an example where one might be able to compare 
between a uMLIP which is aware of the solution and one which is
not, we construct a MACE-MPA $\Delta$-model from the structures gathered
by a REX search for Ag(100)$-(\sqrt{17}\times\sqrt{17})$-Ag$_{12}$S$_8$.
We select the search which contained the lowest energy structure, 
and train the $\Delta$-model on all 200 structures identified by that search. 
This model has thus been informed of the relevant surface phenomena, and 
serves as a point of comparison to the uncorrected uMLIP. Using this
$\Delta$-model as the starting point for new REX repeats we obtain the
purple curve in the right panel of Fig.\ \ref{fig:surface_structures_search}.
This success curve turns out to be a substantial improvement to that of MACE-MPA without
pretraining (cf.\ purple curve in middle panel of Fig.\ \ref{fig:surface_structures_search}),
reducing the average time to reach the solution by up to a full day, 
with the first repeat to find the solution requiring under 6 hours, compared
the 18 hours of the base uMLIP.
Regardless, the induction time present 
before the majority the repeats identify the correct solution suggests 
that the solution is configurationally difficult to come by, requiring 
a reasonable amount of REX relaxations and swaps to manifest. 
The most important conclusion to draw is, however, that 
the magnitude of the improvement achieved with this 
small amount of extra, highly relevant data, is significant. 
It then seems likely that the data gathered by the uMLIP REX 
is particularly descriptive.

To confirm that the data provided to the MACE-MPA $\Delta$-model was indeed 
efficient at describing the relevant regions of the PES, we repeated the 
search with a GPR trained on the same dataset, but with the simple repulsive
prior instead of the uMLIP. The green curve in the last panel of Fig.\ \ref{fig:surface_structures_search} shows the 
success curve of this model, which having seen both the solution and 
a significant amount of relevant data performs as well as the 
corrected MACE-MPA from the previous analysis. This is not 
altogether surprising. It stands to reason that a naive model without 
extensive extrapolation can produce highly accurate results within the interpolation 
domain of its training data. This is compelling evidence that the data
collected by this method of active learning captures the relevant PES 
almost completely. With this we underscore that REX results in both robust 
ab initio quality global optimization and comprehensive sampling of the PES.

\section{Conclusion}

In this article, we have presented universal potential enhanced structure searching, comparing and 
contrasting several methods across a variety of silver sulfide nanocluster stoichiometries and 
surface reconstructions. 

Active learning appears a stable method for gathering data for improving 
the behaviour of uMLIPs. From the four search algorithms considered as 
vehicles for active learning, REX is the clear standout.
We demonstrate significant advantages in efficiency and speed that 
can be achieved when searching with the REX methodology, combining universal potential $\Delta$-models 
with replica exchange search. Regardless of the underlying knowledge (or lack thereof) of a particular 
potential, the REX $\Delta$-model approach is quickly and efficiently able to correct inconsistencies and 
converge to the DFT solution in comparable timeframes. These results encourage that for any universal
potential, such an explorative strategy provides a robust and reliable method to (a) safely apply and (b)
improve and discover, with such potentials. Our findings heavily encourage the adoption of such 
potentials into structure search methodologies, under the provision that they can be corrected 
incrementally with active learning. Universal potential $\Delta$-models resulting from such searches, 
conversely, prove to become more effective as search landscapes than the unaugmented potentials, 
confirming that iterative adaptations effectively alter the form of the PES.

\section{Data availability}
The datasets supporting the findings in this paper are available on Zenodo at 
\url{https://doi.org/10.5281/zenodo.16368472} along with Python scripts to reproduce our findings. The optimization 
methods employed are available in version 3.10.2 of the AGOX package available to install from Pypi 
and source code is openly available on Gitlab \url{https://gitlab.com/agox/agox}.

\section{Acknowledgements}
We acknowledge support from VILLUM FONDEN through Investigator grant, project 
no. 16562, and by the Danish National Research Foundation through the Center of 
Excellence “InterCat” (Grant agreement no: DNRF150).

\bibliographystyle{apsrev4-1}
\bibliography{references}

\end{document}